# Energy dissipation in spintronic digital switches: A general perspective

**Supriyo Bandyopadhyay**
Department of Electrical and Computer Engineering
Virginia Commonwealth University
Richmond, VA 23284, USA

**Abstract:** Champions of "spintronics" often claim that spin based signal processing devices will vastly increase speed and/or reduce power dissipation compared to traditional 'charge based' electronic devices. Yet, not a single spintronic device exists today that can lend credence to this claim. Here, I show that no spintronic device that clones conventional electronic devices, such as field effect transistors and bipolar junction transistors, is likely to reduce power dissipation significantly. For that to happen, spin-based devices must forsake the transistor paradigm of switching states by physical movement of charges, and instead, switch states by flipping spins of *stationary* charges. An embodiment of this approach is the "single spin logic" idea proposed more than 10 years ago. Here, I revisit that idea and present estimates of the switching speed and power dissipation. I show that the Single Spin Switch is far superior to the Spin Field Effect Transistor (or any of its clones) in terms of power dissipation. I also introduce the notion of "matrix element engineering" which will allow one to switch a binary "switch" without raising or lowering energy barriers between the two states of the switch, thereby reducing energy dissipation. Finally, I briefly discuss single spin implementations of classical reversible (conservative) logic.

**Keywords:** Spintronics, single spin logic, power dissipation, reversible logic

## CONTENTS

## 1. INTRODUCTION

A binary switch is the primitive unit of every classical digital computer or signal processor. It processes digital information or signals by switching from one state to another in response to a digital input. It is a bistable component, one of whose states – say, the "on" state - encodes the logic bit 1 and the other – the "off" state – encodes the logic bit 0. The best known electronic switch is a transistor. In the case of a "field effect transistor" (FET) or a "bipolar junction transistor" (BJT), the device is "on" when the active region (the channel of an FET or the base of a BJT) contains a large amount of mobile charges, and it is "off" when that region is depleted of mobile charges. Therefore, switching between logic bits can only be accomplished by *physically moving charges* in and out of the active region with an external agency (such as the gate voltage in an FET or the base current in a BJT). This physical motion consumes considerable energy, which is ultimately dissipated as heat.

The obvious way to reduce the dynamic dissipation during the switching event is to switch between states *without* moving charges. Unfortunately, this is virtually impossible in *charge based* electronics, where the difference in the amount of charge in the active region is used to demarcate logic levels. Charge is a *scalar* quantity and therefore it only has a magnitude. Thus, logic levels can be demarcated solely by a difference in the *magnitude* of the charge, or by the *presence* and *absence* of charge[1]. Therefore, in order to switch from one logic state to another, we must invariably move charges from one region of space to another, thereby causing current flow (*I*) and associated energy dissipation ($I^2R\Delta t = Q^2R/\Delta t$), where $Q$ is the amount of charge moved, $R$ is the resistance in the path of the current and $\Delta t$ is the switching delay. This energy dissipation is unavoidable and it is a fundamental shortcoming of *charge based* electronics.

## 2. SPINTRONICS

Spin, unlike charge, is not a scalar. It has a magnitude and a "polarization". It is easy to make the spin polarization a bistable quantity – and therefore use it to encode binary bits - by simply placing the electron (or hole) in a magnetic field. The Hamiltonian describing a single electron in a magnetic field is

$$H = (\vec{p} - q\vec{A})^2/2m^* - (g/2)\mu_B \vec{B}\cdot\vec{\sigma} \quad (1)$$

where $\vec{A}$ is the vector potential due to the magnetic flux density $\vec{B}$, $\mu_B$ is the Bohr magneton, $g$ is the Lande g-factor, and $\vec{\sigma}$ is the Pauli spin matrix. If the magnetic field is directed in the z-direction ($\vec{B} = B\hat{z}$), then diagonalization of the above Hamiltonian immediately produces two mutually orthogonal eigenspinors [1, 0] and [0, 1] which are the +z and −z-polarized spins, i.e. states whose spin quantization axes are parallel and anti-parallel to the z-directed magnetic field. Thus, the spin quantization axis (or spin polarization) can become a *binary* variable. The "down" (parallel) or "up" (anti-parallel) states can encode logic bits 0 and 1, respectively. These two states are not degenerate in energy, but that does not pose any problem with encoding logic bits.

We could switch from logic bit 0 to 1, or vice versa, by simply flipping spin with an external agent such as a localized magnetic field[2], *without* having to physically move charge and causing current flow. This should eliminate the $I^2R\Delta t$ dissipation. Some energy would still be dissipated since the two states are not degenerate but separated by an amount $g\mu_B B$, but this could be made arbitrarily small by making the magnetic flux density $B$ arbitrarily small. Furthermore, since there is no motion, the switching time is not limited by the transit time of charges, or the velocity of electrons, but instead is limited by the spin flip time which can be engineered to some extent (e.g. by introducing magnetic impurities in the vicinity of the spin). Spin also has another advantage; it does not couple easily to stray electric fields unless the host material has strong spin-orbit interaction. Thus, it has some natural noise immunity, unlike charge.

How stable is the spin polarization? In InAs quantum dots, the single electron spin flip time has been measured to be about 15 ns at 10 K [1]. At room temperature, we expect the spin-flip time to decrease considerably because of spin-phonon coupling [2]. There is currently no published measurement of spin flip times in quantum dots at room temperature. We can - perhaps somewhat optimistically - estimate it to be about 1 ns in InAs quantum dots. If we assume a clock speed of 50 GHz (we will show later that this is a reasonable estimate), then the clock period is 20 ps, which is 1/50-th of the spin flip time. Therefore, the probability that an unintentional spin flip will occur between two successive clock pulses is $1-e^{-1/50}$ = 0.02, or 2%, which can be handled by modern error correction algorithms [3]. Handling an error rate this high requires considerable hardware overhead, but because the device density can be extremely large, this may not turn out to be taxing.

---

[1] It does not have to be "absolute" presence or absence. It has to be simply a "relative" presence or absence. For example 10 electrons could represent 'logic 1' and 2 electrons could represent 'logic 0'. All we need is a 'difference' in the quantity of charge to demarcate logic levels. But in order to switch from one logic state to another, we must be able to alter this difference of 8 electrons, and therefore we must move 8 electrons in space. That consumes energy.

[2] We could also simply reverse the magnetic field to flip spin.

If instead of InAs we choose silicon or InP, then the spin flip time can increases dramatically. Spin flip times of several microseconds have been measured for electrons bound to phosphorus donor atoms in silicon at 20 K [4] and spin flip times exceeding 100 μs have been reported for InP dots at a temperature of 2 K [5]. Room temperature data is unavailable, but even if we assume that the spin flip time is 100 ns at room temperature in silicon or InP quantum dots, the probability of having a random bit flip between two successive clock cycles is $1-e^{-1/5000}$ = 0.02% for 50 GHz clock rate, which could be handled by error correction schemes without mammoth overhead.

## 3. READING AND WRITING SPIN

Encoding binary logic bit information in single electron spin polarization is an interesting notion, but how does one "read" or "write" spin to manipulate or extract this information? In order to controllably orient and detect spin polarization of a single electron (i.e. read and write a bit), we have to first place it in an appropriately designed quantum dot and then place it in a magnetic field that defines the spin quantization axis. The quantum dot will be delineated electrostatically by split metal gates in a penta-layered structure consisting of a ferromagnet-insulator-semiconductor-insulator-ferromagnet combination as shown in Fig. 1(a). The wrap-around split Schottky gate is used for the "write" operation. The ferromagnetic layers also play a critical role in performing the "reading" and "writing" operations.

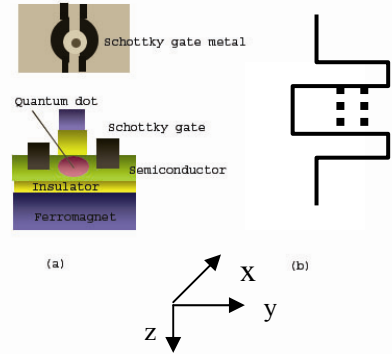

*Fig. 1 (a) Structure of a gated quantum dot to host a single electron with a well defined spin orientation. The top figure shows the top view (in the x-y plane) and the bottom figure shows the cross-section (in the y-z plane). (b) The idealized conduction band energy diagram along the z-direction, perpendicular to the heterointerfaces.*

### 3.1 Writing spin

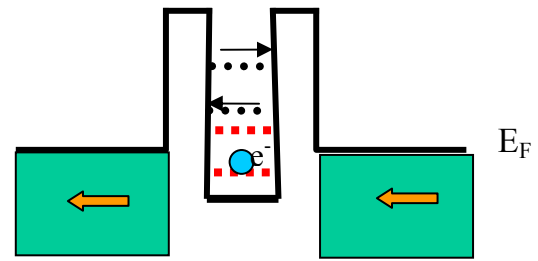

*Fig. 2: The scheme for writing a spin bit. The solid black lines are the conduction band profile in the z-direction before applying the potential to the split gates to enlarge the dot. The dot is "enlarged" in the x-y plane. The broken purple lines are the conduction band profile in the z-direction after enlarging the dot. The lowest spin split state falls below the Fermi level and a single electron occupies the dot.*

The ferromagnet, insulator and semiconductor materials are so chosen that the conduction band energy diagram at equilibrium (in the direction normal to the heterointerfaces) is as shown in Fig. 1(b). The lowest subband (broken lines) is spin split because of the magnetic field caused by the ferromagnetic contacts, plus any other external magnetic field. We will place the Fermi level *below* both spin split levels by appropriate choice of materials and doping. In that case, there will be no electron in the quantum dot (for this discussion, we will assume that the temperature is ~0 K).

In order to "write" a spin bit, a positive potential is applied to the wrap-around gate which decreases the confinement and effectively makes the semiconductor dot larger, thereby pulling the lower spin-split level of the first subband *below* the Fermi level, while still keeping the higher level *above* the Fermi level. This is shown by the broken red lines in Fig. 2. A *single* electron now tunnels into the semiconductor dot from a ferromagnet. Because of Pauli Exclusion Principle, only a single electron can be hosted in the dot. This electron's spin is polarized along the magnetization of the ferromagnet and is therefore known *apriori*. Thus, we have successfully "written" a bit, or a pre-determined spin polarization. Let us say that this bit is logic '0'.

Now, in order to guarantee that the injected spin is always polarized along the direction of the ferromagnet's magnetization, we will really need a ferromagnet with 100% spin polarization. Of course, no known ferromagnet has a 100% spin polarization at room temperature, but it is possible to have rather high spin injection efficiency, as high as 70% at room temperature, even from non-ideal ferromagnets [6]. Thus, there can be ~ 30% error in writing a bit. This error can be corrected by introducing redundancy, i.e. writing the same bit in 3 or more cells and then using majority voting logic to determine the correct bit (a threefold redundancy reduces the error probability to 9% and a fivefold redundancy reduces it to 2.7%). It should be noted that the writing error is a *one-time* error that does not change randomly with time. Such errors are easier to correct than dynamic errors.

The two insulating layers shown in Fig. 1 are important in this context. First, they provide the confinement of carriers in the semiconductor dot. Second, they increase the efficiency of spin injection from the ferromagnet into the semiconductor [7], thereby reducing the error probability during the write operation.

So far, we have described how to write the logic bit 0. But what if we wanted to write logic bit '1' instead? One might guess that we should simply reverse the magnetization of the contacts to do the trick. That would be fine except reversing the magnetization is not so easy. We could not have done it with a magnetic field since the latter would be difficult to confine to a single dot. Therefore, we have to find a different method.

We will do is apply a *differential* potential between the two split Schottky gates in Fig. 1 to induce a Rashba interaction in the chosen dot [8]. The total spin-splitting energy $\Delta$ in the semiconductor layer will then be a mixture of the Rashba and the Zeeman spin splitting. It will be given by approximately [9]

$$\Delta = 2[(g\mu_B B/2)^2 + E_y^2 \frac{16q^2\hbar^4}{m^{*4}c^4 W_x W'_x} f(W_x, W'_x)]^{1/2} \quad (2)$$

where the function $f(W_x, W'_x)$ is

$$f(W_x, W'_x) = \cos^2(\pi W_x / 2W'_x) \cos^2(\pi W'_x / 2W_x) F$$

$$F = \frac{1}{((W_x/W'_x)^2 - 1)^2 ((W'_x/W_x)^2 - 1))^2}$$

Here $E_y$ is the y-directed electric field due to the differential potential applied between the Schottky gates, $W_x$ is the spatial width of the wavefunction along the *x*-direction in the lower Zeeman split spin state, and $W'_x$ is the spatial width in the upper spin state (they are different because the potential barriers confining the electron are of finite height)[3].

---

[3] This expression is based on a simplified treatment that ignores spatial variation of the spin orientation within the dot. A more accurate theory yields a significantly different quantitative result, but is avoided here since it does not yield analytical expressions [see D. Bhowmik and S. Bandyopadhyay, Physica E, 41, 587 (2009) for the

By adjusting $E_y$, we can make the total spin splitting energy $\Delta$ in the chosen dot resonant with a global ac magnetic field of frequency $\omega$ ($\Delta = \hbar\omega$). We hold this resonance for a time duration $\tau$ such that

$$\mu_B B_{ac} \tau = h/2 \quad (3)$$

where $B_{ac}$ is the amplitude of the global ac magnetic field. This is a so-called $\pi$ pulse. It flips the spin and will write the logic bit 1 *only in the chosen dot.* This is the well-known principle used in electron spin resonance. Thus, we can write the logic bit 1 as well.

The magnitude of $E_y$ is of the order of 100 kV/cm. Since the split-gate separation is typically ~ 1 μm, the differential voltage that we have to apply between the two split gates is about 10 V.

We can estimate the value of $\tau$ by assuming a reasonable value of $B_{ac}$ = 0.01 Tesla. In that case, $\tau$ = 18 ps. This tells us what clock speed we can have. The clock speed is limited to 1/(18 ps) = 55 GHz, which is plenty fast.

A much simpler method for reversing the magnetization of nanoscale ferromagnetic contacts is the use of a spin polarized current to exert a *spin transfer torque* on the magnetization vector[4]. This method was not popular at the time the original manuscript was written, but since then has become immensely popular. However, this method dissipates a lot of energy unlike the previous method and is also much slower with $\tau$ ~ 1 ns.

### 3.2 Reading spin

"Reading" a spin, or ascertaining its polarization, is more difficult than writing spin. Single spin reading has been demonstrated with a variety of techniques [10], but most of them are difficult and slow. For electrical detection, we can use the technique of ref. [11] (which is

---

more accurate treatment. This paper appeared two years after this manuscript was originally written.]
[4] For a review, see D. C. Ralph and M. D. Stiles, J. Magn. Magn. Mater., 320, 1190 (2008).

by no means unique and variations exist). This scheme requires the use of ferromagnetic contacts to determine the spin polarization of a target electron. Hence, the need for ferromagnetic contacts.

## 4. SPIN TRANSISTORS

We have described the essential ingredients of a *Single Spin Switch* which acts as a binary logic device. Early research in spintronics however was not concerned with encoding data in single spin polarization. Instead, it concentrated on devising spin based analogs of conventional field effect transistors [12] and bipolar junction transistors [13].

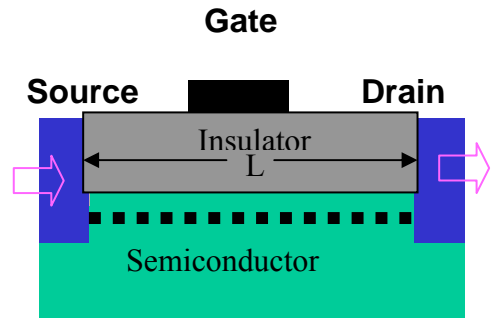

*Fig. 3: Schematic representation of a Spin Field Effect Transistor after ref. [12]. The source and drain contacts are ferromagnets that inject and detect spin of a particular polarization. The gate terminal applies a transverse electric field on the transistor channel (shown with a broken line) which induces a Rashba spin orbit interaction that precesses the spins at a rate determined by the magnitude of the gate voltage.*

Fig. 3 shows a schematic representation of the Spin Field Effect Transistor (SPINFET) proposed in the seminal work of ref. [12]. We will assume that the channel is strictly one-dimensional and only the lowest subband is occupied. This device looks exactly like a conventional metal-oxide-semiconductor field effect transistor (MOSFET) except that the source and drain contacts are ferromagnetic. The ferromagnetic source injects carriers with a particular spin polarization (the majority spins in the ferromagnet) into the channel. The gate voltage induces a Rashba spin-orbit interaction which precesses the spins at a rate that depends on the magnitude of the gate voltage and the carrier momentum. The precession rate in time depends on the gate voltage and carrier momentum, but the precession rate in space can be shown to be independent of the carrier momentum and depend only on the gate voltage. If the gate voltage is such that the spins have precessed by an angle that is an odd multiple of π when they arrive at the drain, then they have a polarization anti-parallel to the drain's magnetization. These spins are blocked by the drain and therefore the source to drain current falls to zero. Without a gate voltage, the spins do not precess[5] and the source to drain current is non-zero. Thus, the gate voltage causes current modulation and realizes transistor action.

It should be obvious that in this device, "spin" actually plays a relatively minor role. Information is still encoded in charge which carries the current. The transistor is still switched by moving charges in space (because a current flows from source to drain). The role of spin is only to provide an alternate means of controlling the current. In ref. [14], we showed that this device does not yield any significant advantage over a conventional charge based transistor such as the MOSFET, unless materials with extremely strong spin-orbit interaction become available.

We showed in ref. [14] that the ratio of the switching voltage of a SPINFET to that of a comparable MOSFET is given by

$$\frac{V_{SPINFET}}{V_{MOSFET}} = \frac{h^2 \pi e}{2m^* L \xi E_F} \quad (4)$$

where $L$ is the channel length, $m^*$ is the carrier effective mass, $e$ is the electronic charge, $h$ is the Planck constant, $E_F$ is the Fermi energy in the channel, and $\xi$ is the change in the Rashba spin orbit interaction strength per unit change in the gate voltage. The quantity $\xi$ has been measured in InAs and reported in ref. [15]. If we assume the measured value, then the ratio in Equation (4) will be less than unity only if the channel length exceeds 295 μm in an one-dimensional InAs SPINFET with carrier concentration (and therefore $E_F$) small enough to maintain single subband occupancy. In other words, the SPINFET can have a lower switching voltage (and therefore lower power dissipation) than the MOSFET only if it is impractically long. On the other hand, if we assume the theoretical maximum value of $\xi$, then the SPINFET has a lower threshold voltage if the channel length is larger than 4.9 μm. Thus, no sub-micron SPINFET, let alone a nanometer scale SPINFET, is likely to yield any advantage over a comparable MOSFET. This is because we are still switching the SPINFET by moving charges in space. As long as we are within that paradigm, we cannot expect any significant improvement in power dissipation over charge based devices.

Many proposals have now appeared in the literature claiming to improve the design of the original SPINFET of ref. [12]. One of them [16] posits a device that can operate in the diffusive regime unlike the SPINFET of ref. [12] which normally requires ballistic transport (see footnote [2]). In this device, 100% spin polarized current is assumed to be injected from the ferromagnetic source contact. In the absence of any gate voltage, there is no spin flip scattering in the channel so that the injected spins arrive intact at the ferromagnetic drain. The drain (which is magnetized parallel to the source), transmits all of these spins and the current is maximum. When the gate voltage is turned on, it changes the Rashba spin orbit interaction in the channel which causes spin flip scatterings. As a result, some carriers flip spin. The carriers with flipped spins are blocked by the drain. At best 50% of the spins will be flipped, at which point, the spin polarization of the current becomes zero. Thus the current can drop to one-half of the maximum value when the gate voltage is turned on.

It is obvious that this device can provide a maximum ratio of the "on"-conductance to the "off"-conductance = 2, instead of the 1000

---

[5] Even without the gate voltage there is obviously some Rashba interaction in the channel since the structure is not symmetric. We ignore that here.

required for most applications[6]. This ratio can improve if the device is modified so that the two ferromagnetic contacts are magnetized in anti-parallel directions, unlike the device of Fig. 3 where the two contacts are magnetized in parallel direction. If nearly perfect spin injection from the ferromagnetic contacts can be attained, the anti-parallel arrangement can cause dramatic improvement in the conductance ratio. In fact, 100% spin injection and spin extraction efficiencies would make the conductance ratio infinity. However, perfect spin injectors or spin detectors do not exist as yet and the maximum spin injection efficiency demonstrated so far is 90% from non-permanent ferromagnets [18] and 70% from permanent ferromagnets [6]. In the former case, 5% of the total injected current is due to minority spins, and in the latter case, 15% is due to minority spins. Consequently, even if the drain contact is a perfect spin detector, the leakage current flowing during the "off" state will be 5% or 15% of the injected current. The current flowing during the "on" state is 50% of the injected current at best (when the spin flip scattering completely destroys any net spin polarization in the current). Therefore, the on-to-off conductance ratio is 0.5/0.05 = 10, or 0.5/0.15 = 3.33, in the two cases. That is not a significant improvement over the factor of 2 attained when the two contacts have parallel magnetizations.

Anti-parallel magnetizations however reduce the stray magnetic field in the channel of a SPINFET. This can improve the performance of the SPINEFT of ref. [12] in a number of ways, as discussed in refs. [19] and [20].

Other SPINFET proposals have recently appeared in the literature. One of them [21] is similar to the proposal of ref. [16] in that the gate voltage changes the spin flip scattering rate, and thus modulates the source-to-drain current. Because of this similarity, it inevitably suffers from the same drawback as ref. [16], namely small on-to-off conductance ratio, even for anti-parallel magnetizations, as long as we assume realistic spin injection efficiencies. However, ref. [21] makes an important claim. It claims that "[this] spin transistor operation will be possible at a much lower threshold voltage than conventional CMOS technology". Ref. [22] expands on this claim and asserts that the SPINFET will have a lower leakage current in the OFF-state than a MOSFET. We have recently shown that these claims are invalid [23]. The apparent basis of such claims is the authors' belief that a small voltage can induce a large modulation of the Rashba interaction in the channel of quantum wells, so that a SPINFET could be switched with a smaller gate voltage than a MOSFET. This claim is in direct contradiction with experimental findings of ref. [15] which shows that the gate modulation of the Rashba interaction is actually rather weak. The quantity $\xi$ was experimentally measured to be only $8 \times 10^{-31}$ C m and even the maximum theoretical value is only 60 times larger [14].

---

[6] Simulations [17] show that the actual ratio is even less than 2; it is only about 1.2. Therefore, it is clearly insufficient for device applications.

Ref. [21] calculates that in a channel with Fermi energy $E_F$ = 30 meV, a gate voltage of 140 mV should reduce the spin lifetime to ~ 10 ps. It will be reasonable to assume that the carrier mobility at room temperature will be ~ 1 $m^2$/V-sec in the InAs channel considered. Then, the spin diffusion length will be ~ 0.16 μm at a gate voltage of 140 mV. Thus, a device with channel length $\geq$ 0.16 μm can be turned on and off with a gate voltage of 140 mV. A shorter device will require a larger gate voltage.

Based on the above, the switching voltage of a 0.16 μm long SPINFET can be only 140 mV. That may be true, but if this *same device* were used as an ideal MOSFET instead of an ideal SPINFET, the voltage required to deplete the channel completely (and therefore switch the transistor) would have been simply $E_F/e$ = 30 mV. Therefore, this spin device will not have a lower switching voltage, but rather a higher switching voltage by a factor of 140/30 = 4.6. Note that in calculating the switching voltage, we have consistently neglected any voltage drop across the gate insulator, but that does not change our conclusion.

In ref. [23], we also showed that this type of SPINFET has a particularly large leakage current because of the small ratio of the on-to-off conductance. The leakage current is consequence of imperfect spin injection and detection. We showed that in order to have the off-current smaller than 0.1% of the on-current, one would need the spin injection efficiency at the ferromagnet/semiconductor interface to exceed 99.9%. That is a very tall order.

We have also analyzed the spin bipolar junction transistors proposed in ref. [13] and found that they too are *not* significantly better than conventional (charge based) bipolar junction transistors [24]. Again, the reason is that "spin" plays a minor role in their operation and switching is still accomplished by moving charges in space.

## 5. SINGLE SPIN LOGIC FAMILY

Significantly lower threshold voltages can only be achieved via a paradigm shift, namely by switching states without physically moving charges. Consistent with this viewpoint, we proposed the idea of "single spin logic" in 1994 where a single electron acts as a binary switch and its two orthogonal (anti-parallel) spin polarizations encode binary bits 0 and 1 [25]. Switching is accomplished by flipping the spin without moving charges. This is the first known logic family based on single electron spins.

Recent advances in controlling and manipulating an electron at the single spin level [26-34] has now made it possible to make important strides towards the implementation of single spin logic. This motivates the present review.

Many logic circuits, both combinational and sequential, have been designed and theoretically verified using the single spin idea [25, 35]. Here, I will repeat the design of a NAND gate since it is a universal gate. Any logic circuit can be realized with it.

### 5.1. The NAND gate

Consider a linear chain of three electrons in three quantum dots as shown in Fig. 4. They are placed in a global (dc) magnetic field, as mentioned before, to make the spin polarization a bistable entity. We will assume that only nearest neighbor electrons interact via exchange since their wavefunctions overlap. Second nearest neighbor interactions are negligible since exchange interaction decays exponentially with distance.

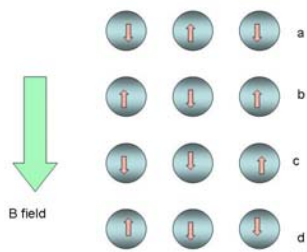

*Fig. 4: Spin configurations in a 3-dot system with nearest neighbor exchange coupling. The peripheral spins are the two inputs and the central spin is the output. The four configurations correspond to the four entries in the truth table of a NAND gate. The B-field is necessary to make spin polarization a bistable entity and to resolve "ties".*

We will also assume that the exchange energy (defined as the energy difference between the triplet and singlet states of two neighboring interacting electrons) is larger than the Zeeman splitting energy caused by the globally applied external (dc) magnetic field. In this case, the ground state of the linear array has *anti-ferromagnetic ordering* where nearest neighbors have opposite spin. In fact, the ground state of the array looks like as in Fig. 4(a). This was verified by quantum mechanical many-body calculation in ref. [36].

Let us now regard the two peripheral spins as the two "inputs" to the logic gate and the central spin as the "output". Assume the downspin state $(\downarrow)$ [parallel to the global magnetic field] represents logic 1, and the upspin state $(\uparrow)$ is logic 0. Then, we find that when the two inputs are 0, the output is automatically 1 because the ground state has anti-ferromagnetic ordering. That is encouraging since it is one of the four entries in the truth table of a NAND gate.

We now have to realize the other three entries in the truth table. If we change the two inputs to 1 from 0, using the technique embodied in Equation (3), then this takes the system to an excited state since the ordering is no longer anti-ferromagnetic. We then let the system relax. In order to regain anti-ferromagnetic ordering, either the central spin will flip down (e.g. by emitting a phonon), or else the two peripheral spins will flip down to their original state after the writing operation is complete. The former requires a *single* spin flip, while the latter requires *two* spin flips. However, the former process will take the system to a local metastable state, while the latter takes it to the global ground state. Note that the states in Figs. 4(a) and 4(b) are not degenerate in energy because of the global magnetic field, which favors 4(a) over 4(b).

Whether the metastable state is reached, or the global ground state is reached depends on the energy landscape and the switching dynamics. If the time $\tau$ in Equation (3) is much smaller than the spontaneous spin flip time, then we will likely reach the metastable state. If the metastable state is reached, then we will achieve the configuration shown in Fig. 4(b). This is the desired configuration because when the two inputs are 1, the output is 0. This is yet another entry in the truth table of a NAND gate. The metastable state must ultimately decay to the global ground state by flipping spins spontaneously. But since the spin flip time is typically much longer than the inverse of the input data rate (~ 50 GHz), we can ignore it.

Finally, what happens if one input is 1 and the other 0? This situation seemingly causes a tie, but the global magnetic field resolves this situation in favor of the central dot having a down-spin configuration (parallel to the magnetic field). The corresponding spin arrangements are shown in Figs. 4(c) and 4(d). Note that these conform to the other two entries in the truth table of a NAND gate.

Finally, we have realized the entire truth table:

| Input 1 | Input 2 | Output |
|---------|---------|--------|
| 0 | 0 | 1 |
| 0 | 1 | 1 |
| 1 | 0 | 1 |
| 1 | 1 | 0 |

In 1995 Molotkov and co-workers verified the entire truth table by carrying out a fully quantum mechanical calculation of the various spin configurations in a 3-dot system [36]. Further work in this area has been performed by Bychkov and co-workers [37].

Note that the global magnetic field serves dual purposes: (i) it defines the spin quantization axis while making spin a binary variable, and (ii) it resolves 'ties' when they arise. Without the global magnetic field, this paradigm would not work.

A similar idea for implementing logic gates using single electron charges confined in 'quantum dashes' was proposed by Pradeep Bakshi and co-workers in 1991 [38]. There, logic bits were encoded in bistable charge polarizations of elongated quantum dots known as 'quantum dashes'. Coulomb interaction between nearest neighbor quantum dashes makes the ground state charge configuration "anti-ferroelectric" just as exchange interaction in our case makes the ground state spin configuration anti-ferromagnetic. Three Coulomb coupled quantum dashes would realize a NAND gate in a way very similar to what was described here.

Bakshi's idea inspired a closely related idea known as 'quantum cellular automata' [39]

which uses a slightly different host, namely a 4- or 5-quantum dot "cell" instead of a quantum dash to store a bit. Here too the charge polarization of the cell is bistable and encodes the two logic bits. Coulomb interaction makes the ground state charge configuration ferroelectric. Logic gates are implemented in the usual fashion.

The paradigm of ref. [38] and its clone [39] involve *charge motion* for switching. Thus, they are likely to be more dissipative than single spin logic which eliminates charge motion altogether. But more importantly, it is very difficult to enforce only nearest neighbor interactions because the Coulomb interaction is long range. If second nearest neighbor interactions are not very much weaker than the nearest neighbor ones, the ground state charge configuration is only marginally stable and not robust against noise. In this respect, our spin based approach has an advantage. Since exchange interaction is short range, it is much easier to make the second nearest neighbor interactions considerably weaker than the nearest neighbor interactions. Accordingly, the anti-ferromagnetic state is much more stable than the anti-ferroelectric (or ferroelectric) state against other spurious states. This, coupled with the fact that spin does not easily couple to stray electric fields unlike charge, makes the single spin logic gates superior.

**5.2 The issue of unidirectionality**

There is, however, a serious problem with these types of logic gates which may not be apparent at first. There is no *isolation* between the input and output of the logic gate since exchange interaction is "bidirectional". Consider just two exchange coupled spins in two neighboring quantum dots. They will form a singlet state and therefore act as a natural NOT gate. However, exchange interaction cannot distinguish between which spin is the input bit and which is the output. Since the input and output are indistinguishable, it becomes ultimately impossible for logic signal to flow *unidirectionally* from an input stage to an output stage and not the other way around. We have discussed this issue at length in various publications [25, 40] since it is vital. Because of this problem, the ideas of ref. [38] and [39], as proposed, could not work since Coulomb interaction is also bidirectional.

It is of course possible to forcibly impose unidirectionality by holding the input cell in a fixed state until the desired output is produced. In that case, the input signal itself enforces unidirectionality because it is a symmetry-breaking influence. This approach was actually used to demonstrate a 'magnetic cellular automaton' where the input enforced unidirectionality and produced the correct output [41]. However, there are problems. First, this approach can only work for a small number of cells before the influence of the input dies out. Second, and more important, the input cannot be changed until the final output has been produced since otherwise the correct output may not be produced at all. That makes such architectures *non-pipelined* and therefore unacceptably slow. There may also be additional problems associated with random errors when this approach is employed. They have been discussed in ref. [42].

In 1994, when we first proposed the single spin logic gates, we thought of enforcing unidirectionality artificially by progressively increasing the distance between quantum dots, so that there is spatial symmetry breaking [25]. In 1996, we revised this idea and instead pointed out the possibility of imposing unidirectional flow of signal in *time*, rather than in *space*, by using clocking to activate successive stages sequentially in time [43]. The actual clocking can be done in the same way as is done in bucket brigade devices, such as charge-coupled-device (CCD) shift registers[7], where a push clock and a drop clock are used to lower and raise barriers and thus steer a charge packet unidirectionally from one device to the next. In single spin circuits, during the positive cycle, the clock signal will apply a positive potential to the barrier separating two neighboring quantum dots which will lower the barrier temporarily to exchange couple these two spins. This will result in the two spins assuming anti-parallel polarizations. Then during the negative clock cycle, the barrier is raised again to decouple the two spins. In this fashion, spin bits can be transferred unidirectionally from one dot to the next. Just as in CCD devices, a single phase clock cannot impose the required unidirectionality; a 3-phase clock is required to do this job [44].

---

[7] CCDs also have no inherent unidirectionality. There a push clock and a drop clock are used to steer charge packets from one device to the next. See, for example, D. K. Schroeder, *Advanced MOS Devices*, Modular Series in Solid State Devices, Eds. G. W. Neudeck and R. F. Pierret (Addison-Wesley, Reading, MA, 1987).

The clocking circuit however introduces additional dissipation. More importantly, it takes away the most attractive feature of these circuits, namely the *absence* of interconnects (or "wires") between successive devices. In Single Spin Logic, exchange interaction plays the role of physical wires to transmit signal between neighboring spins (switches), but in order to steer bits "unidirectionally" down a logic chain, we will need each switch or spin to be clocked individually and that requires a physical interconnect (clock pad) to be placed between every pair of quantum dots. The split gates in Fig. 1 can be used as the clock pads, but wires must be attached to them to ferry the clock signal. Therefore, the architecture is no longer wireless. More importantly, the lithographic burden is daunting. Neighboring dots should not be separated by more than a few nm in order to retain adequate exchange coupling and this mandates interposing a gate pad within a very narrow space (~few nm), which is extremely challenging. All this of course detracts from the appeal of a "wireless architecture", or the so-called "quantum coupled architecture", but this is the price one must pay when inter-device communication is via bidirectional exchange interaction.

### 5.3 Energy and power dissipation

We are now ready to calculate the energy and power dissipation per bit flip in Single Spin Logic. When a bit flips, the energy released (or absorbed) is the energy difference $\Delta E$ between the two spin states in a quantum dot which will depend on the spin orientations of its neighbors because of exchange interaction. Roughly speaking, this energy will be of the order of the exchange splitting energy. Reasonable upper estimates for the latter in today's coupled dot systems is about 1 meV [45], so that $\Delta E \sim 1$ meV.

Now, if the energy change in transitioning between the logic states is ~1 meV, then the energy dissipated in switching is also no more than 1 meV = $1.61 \times 10^{-22}$ Joules. The corresponding power dissipation for a 50 GHz clock is only 8 pW.

### 5.4 Operating temperature

The reader may wonder at this point what the temperature of operation will be if the energy difference between the two logic states is only 1 meV. The answer depends on whether the spin system is in thermal equilibrium with the surroundings. If it is, then the relative occupation probability of the two spin states will be governed by Boltzmann statistics and the probability of spontaneous occupation of the excited state – which is the error probability – will be $p = \exp\left[\dfrac{-\Delta E}{kT}\right]$. This yields $kT = \dfrac{\Delta E}{\ln(1/p)}$. With $p = 10^{-9}$ [a reasonable error probability], $kT$ will be 47.6 μeV, or $T = 0.55$ K. This would doom us to cryogenic operation, making Single Spin Logic entirely impractical.

However, a spin system does not need to be in thermal equilibrium with its environment. Equilibration takes place only if thermal perturbations can flip spin to make the spin distribution conform to Boltzmann or Fermi-Dirac statistics. In reality, fluctuations alone *cannot* flip spin. Otherwise, electron spin resonance experiments – where the spin splitting energies are usually less than 1 meV – could not be carried out at room temperature. Thermal fluctuations (phonons) may supply the energy required to cause a spin flip, but supplying the energy alone is not sufficient to initiate a spin flip. As ref. [2] points out: "the phonon itself does not flip the spin but [merely] provides energy conservation". Spin flips are caused by spin-orbit coupling, interactions with nuclear spins, magnetic impurities, etc. [2]. While the strengths of these interactions and the rate of spin flip can certainly go up with temperature, there is no connection between the thermal energy and the energy difference between the two spin states in causing unwanted bit flips. In other words, if the spin system does not equilibrate with the thermal bath, then it is very possible to have $kT \gg g\mu_B B$ and yet have a reasonable error probability since $p \neq \exp\left[\dfrac{-g\mu_B B}{kT}\right]$ when the spins are not in equilibrium with the thermal bath. As long as we can maintain the spins out of equilibrium, even room temperature operation is not out of the question.

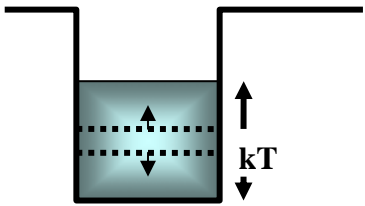

*Fig. 5: The energy splitting between the spin levels may be much smaller than kT. Spin levels are not broadened significantly by coupling to phonons since the spin-phonon coupling is very weak.*

Another issue that often raises doubts is the following: how can one resolve an energy splitting of 1 meV at room temperature? Are not the levels broadened by ~ $kT$ = 25 meV at room temperature which should swamp the levels? The answer is *no*. Spin does not couple strongly to phonons and therefore the level broadening is much less than $kT$ which allows one to resolve states separated by 1 meV at room temperature. If this were not the case, then again electron spin resonance experiments could not be carried out at room temperature. The same weak coupling between spins and phonons also makes it much easier to maintain a spin system out of thermal equilibrium than a charge system since charge-phonon coupling is much stronger. In the end, this is another advantage of spin over charge.

### 5.5 What about the Landauer result?

The reader will also likely have another doubt about room temperature operation. If the operating temperature is 300 K so that $kT$ = 25 meV, then it seems implausible that the energy dissipation to switch can be only 1 meV since that seems to contradict a popular idea which states that the minimum energy dissipated in an irreversible logic operation is *kTln2* [45]. At room temperature, *kTln2* is 18.75 meV, which is nearly 19 times the energy dissipated! So, how is this possible?

In reality, the *kTln2* limit, known as the Landauer limit after its author, is valid only when the switch is in thermal equilibrium with its surroundings. If the switch is far out of equilibrium, then this limit has no relevance and does not apply. Even in thermal equilibrium, the "Landauer limit" is not straightforward and there are subtleties associated with it. Landauer showed that only if we switch in a complicated way with a very specific sequence of clocking [45], we will switch with 100% probability while dissipating *kTln2* amount of energy. But if we switch in a straightforward fashion by simply tilting the potential profile of the switch to bias it towards the desired state, then the minimum energy dissipated in switching is $kT\ln(1/p)$, where *p* is the *static bit error probability*. All this is of course valid only in equilibrium and has no validity for out-of-equilibrium systems.

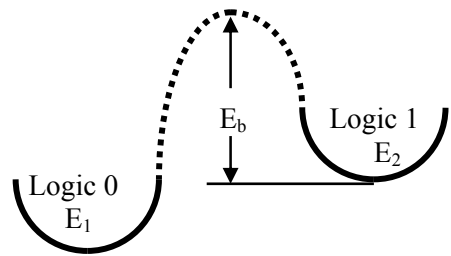

*Fig. 6: Energy landscape for a binary switch*

Let us examine the basis of this result. A binary switch has two distinguishable states that encode logic bits 0 and 1. The potential energy landscape of such a switch is shown in Fig. 6 where the two potential wells represent the two stable states. Note that we have intentionally made the two states non-degenerate in energy ($E_1 \neq E_2$) to correspond to our situation where they are separated by $\Delta E$.

The rate of transition between these two states will be normally given by Fermi's Golden Rule

$$S(E_1, E_2) = \frac{2\pi}{\hbar} | M_{E_1, E_2} |^2 \delta(E_2 - E_1 \pm \hbar\omega) \quad (5)$$

where $M_{E_1,E_2}$ is the matrix element for transition between the states and $\hbar\omega$ is the energy transfer.

Now, we do not want spontaneous transitions to occur when the switch is not activated because that will result in a bit error. Therefore, we must make $S(E_1, E_2)$ zero when the switch is inactive and non-zero when the switch is active.

There are two ways to achieve this. First, we can build an energy barrier between the two states so that one is not *directly* accessible from the other. This is shown by the broken line in Fig. 6. In that case, the only way spontaneous transitions can take place is if there is tunneling through the barrier (which is negligible if the barrier is tall enough or wide enough), or there is thermionic emission over the barrier. If the switch is in thermal equilibrium and the occupation probability of a state is given by Boltzmann statistics, then the probability of spontaneous thermionic emission over a barrier of energy $E_b$ is $p_0 = e^{\frac{-E_b}{kT}}$, which yields $E_b = kT \ln(1/p_0)$. Clearly, thermionic emission will cause spontaneous switching even when the switch is inactive. Hence, the static bit error probability is essentially the probability of thermionic emission (neglecting tunneling). Therefore, $p = p_0$ and we get the result $E_b = kT \ln(1/p)$. This result tells us that at a given temperature, the taller the barrier is the smaller will be the static bit error probability.

In order to active the switch and induce switching, we can tilt the potential profile in Fig. 6 towards the left to increase the energy difference $E_2 - E_1$ if the desired final state is $E_1$ and the initial state is $E_2$. To be safe, we should make $E_2 - E_1$ at least equal to $E_b$, which will ensure that the barrier separating the two states is completely eroded. In that case, the system will switch from state $E_2$ to state $E_1$ with very high probability since there is no barrier impeding the transition. The energy dissipated in switching is then

$$E_2 - E_1 = E_b = kT \ln(1/p).$$

This is of course a "brute-force" method of switching which is more dissipative than necessary. There are more subtle and less dissipative switching strategies but those will entail exquisite timing synchronizations [45] that can introduce large *dynamic* errors during switching.

There is however a second way to prevent unintentional switching when the switch is inactive and induce switching when the switch is activated. This is usually not considered, but we discuss it here. We can make the matrix element in Equation (5) nearly zero when the switch is inactive and large when the switch is active. Thus, we have to *modulate the matrix element*, making it vanishingly small during the inactive phase and large during the active phase. This "matrix element engineering" can replace "barrier engineering" and eliminate the need for the barrier. In other words, *we switch not by tilting the potential profile to erode the barrier, but by modulating the matrix element*. Since there is no tilting by an amount equal to or greater than the barrier height, the minimum energy dissipated during switching need not be $E_b = kT \ln(1/p)$, but can be considerably less, or even zero. We illustrate this with a concrete example.

### 5.6 Matrix element engineering

In order to implement "matrix element engineering" we have to ensure that the matrix element is zero when the switch is inactive and non-zero when active. Let us consider the case of a single spin placed in a magnetic field of flux density $B$. The two allowed spin states are parallel and anti-parallel to the field and are designated as $|\uparrow\rangle$ and $|\downarrow\rangle$. They are separated in energy by $g\mu_B B$.

We will assume that there is no agent (no magnetic impurity, no hyperfine interaction, etc.) that can couple the two mutually anti-parallel spin states in the dot, so that no spin flip transition can occur and the matrix element for spin flip transition is zero. Hence, an excited spin state cannot decay to the ground state by emitting phonons, magnons, etc. In other words, the switch is *inactive*.

In order to activate the switch, we apply a π-pulse with an ac magnetic field whose frequency is *resonant* with the spin splitting, i.e. $\hbar\omega = g\mu_B B$, where $\omega$ is the angular frequency of the field. This will flip the spin. The pulse makes the switch *active* temporarily and makes the matrix element temporarily non-zero (only over the duration of the pulse). Once the pulse is removed, the switch reverts to the inactive state. The pulse therefore implements matrix element engineering; its presence makes the matrix element non-zero and its absence makes the matrix element zero.

Dynamical switching of a spin by this method does not dissipate any energy at all since the spin rotates by coherently exchanging photons with the electromagnetic wave producing the ac magnetic field. It will absorb a photon of energy $\Delta E$ if the final state is $E_2$ and emit a photon of energy $\Delta E$ back to the field if the final state is $E_1$. The only caveat is that we must be exquisitely precise with the pulse period and frequency since there is no tolerance for error here. The point to note is that we can deterministically switch without dissipating an amount of energy equal to $E_2 - E_1 = g\mu_B B = kT \ln(1/p)$. This has been made possible by matrix element engineering.

This method of switching a spin is of course very well known and routinely used in electron spin resonance spectroscopy. We mention it here in the context of switching since it is very different from the traditional switching methods of raising and lowering barriers to switch.

### 5.7 Energy dissipation in the clock

Finally, there is also the issue of how much power is dissipated in the clock that steers bits unidirectionally in Single Spin Logic. For square-wave clock pulses, the energy dissipated in each clock pad is $CV^2$ where $C$ is the capacitance of the pad and $V$ is the voltage applied to lower the potential barrier between neighboring cells. We can of course reduce this energy dramatically if we do not switch abruptly, but switch slowly, but then that would tend to increase clock error rates and reduce bit propagation rate. We can also eliminate this dissipation altogether by using a resonant LRC circuit where the inductor and capacitor are in parallel and the resistance is in series, but this is ultimately cumbersome. Therefore, let us stick with a capacitor and abrupt clock pulse.

If $C$ = 1 aF and $V$ = 100 mV (which are reasonable estimates), the energy dissipation in a clock pad is $10^{-20}$ Joules and the clock power dissipated with a 50 GHz clock frequency = 0.5 nW per pad. Thus, the clock dissipates 625 times more energy than the device itself. Once again, if this is unacceptable, we can always resort to adiabatic clocking and sacrifice bit propagation speed.

Assuming that there are $10^{11}$ quantum dots and therefore $10^{11}$ clock pads/cm$^2$, the total power dissipated in the chip is 50 Watts/cm$^2$ in the worst case, assuming that all spins are flipping simultaneously. With a more reasonable 10% activity level, i.e. assuming only 1 in 10 spins is flipping at any given instant of time, the power dissipation will be 5 Watts/cm$^2$.

The Intel Pentium 4 chip of circa 2000 has a transistor density slightly less than $10^8$/cm$^2$ [48], which is 3 orders of magnitude less than what we assume for single spin based switches. The Pentium IV dissipates about 50 Watts/cm$^2$ [49] with somewhere in the vicinity of 10% activity level. Thus, with Single Spin Logic, one can potentially increase the bit density by 1000 times while keeping the power dissipation per unit area 10 times smaller. All this is of course theoretical speculation since there is no experimental report of anything approaching Single Spin Logic at the time of writing.

**5.8 Comparison with the SPINFET**

We will now compare the Single Spin Switch with the Spin Field Effect Transistor, which, in some ways, is a fair comparison since neither has been demonstrated experimentally. The switching voltage of a SPINFET is $h^2\pi/(2m^*L\xi)$ [14]. In our calculations, we will assume the theoretical maximum value of $\xi$ given in ref. [14], which is 5 x 10$^{-29}$ C m. Since 90 nm gate length CMOS devices are available and since Single Spin Switches can be easily manufactured with feature sizes less than 90 nm, we will assume that the SPINFET has a gate length $L$ = 90 nm for a fair comparison.

In that case, the switching voltage of an InAs SPINFET is $V_{switch}$ = 5.6 V. If we assume that the gate capacitor has a width of 90 nm and the gate insulator has a relative dielectric constant of 4 and a thickness of 10 nm, then the gate capacitance $C_{gate}$ = 28 aF. Accordingly, the energy dissipated in switching = $C_{gate}V^2_{switch}$ = 8.7 x 10$^{-16}$ Joules, which is higher than what present day transistors dissipate [49] and nearly five orders of magnitude higher than what is dissipated in the clock of single spin logic. With a 30 GHz clock, the power dissipated will be 26 μW/bit flip, which is higher than what transistors today dissipate [49].

A nanoscale SPINFET, therefore, is *not* a low-power device.

**6. SPINTRONIC REVERSIBLE (ADIABATIC) LOGIC**

So far, we have discussed a spintronic logic family (Single Spin Logic) that dissipates very little energy during switching. But can we design logic gates that dissipate no energy at all?

It is well known that such gates must be logically reversible [50], i.e. we should be able to infer the input state unambiguously from the output state. There is a vast body of literature on reversible computers, which dissipate no energy at all [50]. Quantum computing is a subset of reversible computing.

In 1996, we proposed an idea to implement a *quantum adiabatic inverter* with single spins [51]. Just two exchange coupled spins, in two closely spaced quantum dots placed in a weak external magnetic field, make a quantum inverter. This gate is logically reversible since the input can always be inferred from the output (they are simply logic complements of each other). The inverter could be switched adiabatically without dissipating any energy at all and ref [51, 52] discuss some interesting results pertaining to the quantum mechanical evolution of this system in time deduced from the time-dependent Schrődinger equation.

We showed that the switching time of the optimally designed adiabatic inverter is

$$t = \frac{h}{8\sqrt{2}J} \quad (6)$$

where $J$ is the exchange energy, or the energy difference between the triplet and singlet configuration of the two spins. If $J$ = 1 meV, $t$ is less than 1 ps. Therefore, this gate switches quite fast and dissipates no energy.

The problem with such adiabatic devices is that they have no fault tolerance and there is a "halting problem". For example, the quantum inverter must be halted after precisely the time duration given in Equation (6), if we want the desired output. Otherwise, the system will continue to evolve with time and the output will continue to drift from the correct state. The system will periodically revisit the correct state since it is reversible and therefore obeys the principle of Poincare recurrence. However, the halting problem is a major inconvenience. These devices are interesting theoretical curiosities, but probably not very practical at this time.

In ref. [51] we also implicitly posited the idea of using the spin of an electron in a quantum dot to represent a "qubit", although we did not use the term "qubit" since it had not been coined yet.

This idea has now become a widely popular concept in the context of quantum computing after Loss and DiVincenzo showed that a *universal* quantum gate can be realized based on two exchange coupled spins [53].

**6.1 Toffoli-Fredkin gate**

The inverter designed in ref. [51] is not a *universal* adiabatic gate, since not any arbitrary adiabatic circuit can be realized with an inverter alone. The universal adiabatic gate is the *Toffoli-Fredkin (T-F) gate* which has three inputs $A,B,C$ and three outputs $A',B',C'$. [54]. Input-output relationships for the T-F gate are $A' = A$, $B' = B$ and $C' = C \oplus A \bullet B$, where $\oplus$ is the EXCLUSIVE-OR operation and $\bullet$ is the AND operation. The T-F gate requires that $C$ toggle if, and only if, $A$ and $B$ are both logic 1; otherwise, nothing should happen.

We can realize this gate with three spins placed in a global magnetic field with nearest neighbor exchange interaction (exactly the same configuration as the NAND gate described earlier). The two peripheral spins are the control bits $A$ and $B$ and the central spin is the bit $C$.

There is only one difference with the NAND gate. This time, we will make the Zeeman interaction (due to the magnetic field) stronger than the exchange interaction.
The Zeeman interaction makes the downspin state lower in energy than the upspin state in each dot. The exchange interaction, on the other hand, tries to make spins in neighboring dots anti-parallel. The interplay of these two effects realizes the T-F gate.

If the spins in $A$ and $B$ are "down", then the exchange energy will tend to keep the spin in $C$ "up", while the stronger Zeeman interaction still retains the downspin state (the state parallel to the magnetic field) as the lower energy state. In this case, the exchange interaction *subtracts* from the Zeeman splitting in the central dot and makes the total spin splitting energy in this dot less than the bare Zeeman splitting. If the spin in dot $A$ is "up" and that in dot $B$ "down", then the exchange interaction effects due to $A$ and $B$ on the spin in dot $C$ tend to *cancel* and the spin splitting in dot $C$ is more or less the Zeeman energy. On the other hand, if the spins $A$ and $B$ are both "up", then the exchange interactions due to them *add* to the Zeeman splitting in dot $C$, making the total spin splitting energy in $C$ larger than the bare Zeeman splitting. In essence, the total spin splitting energy in $C$ is *larger* when both $A$ and $B$ are in logic 1 state, than otherwise.

If we denote the total spin splitting energy in dot $C$ as $\Delta_{\alpha\beta}^C$ where $\alpha$ and $\beta$ are the logic states in dots $A$ and $B$, then the following inequality holds:

$$\Delta_{11}^C \geq \Delta_{10}^C = \Delta_{01}^C \geq \Delta_{00}^C$$

These situations are shown in the energy diagrams in Fig. 7.

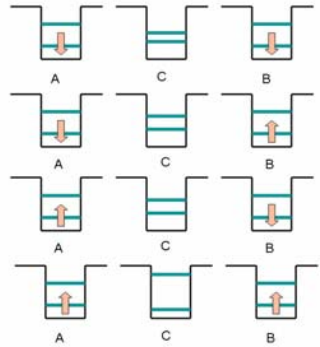

*Fig. 7: The realization of a spintronic Toffoli-Fredkin gate. The spin splitting energies in the central dot are shown as a function of the spin orientations in the two peripheral dots. The two peripheral dots are the two control bits 'A' and 'B' and the central spin is the target bit 'C'.*

To implement the dynamics of the T-F gate, the whole system is pulsed with a global ac magnetic field of amplitude $B_{ac}$ whose frequency is resonant with $\Delta_{11}^C$. The pulse duration is $h/(2\mu_B B_{ac})$. Therefore, $C$ will toggle if $A$ and $B$ are both in logic state 1. Otherwise, nothing will happen. Thus, we have realized the truth table of a Toffoli-Fredkin gate. This is the standard technique for realizing this gate (see ref. [55] for a similar idea). The Toffoli-Fredkin realization leads to an entire family of reversible logic gates based on exchange coupled single spins in a global dc magnetic field. Note that in the case of the T-F gate, we assume that the π-pulse induces *coherent* rotation of the spin in dot C. This is necessary to ensure adiabaticity.

## 7. CONCLUSION

In this work, I have reviewed some examples where the spin polarization of a charge carrier is used to encode bits of information. Signal is processed without physically moving charges, thereby saving energy in the device. These devices are "classical" and therefore spin coherence is not an issue. They are considerably easier to implement than spin-based quantum computing.

I have also introduced the concept of matrix element engineering that can be beneficially employed to switch binary switches with arbitrarily small dissipation. Admittedly this is not easy to implement and requires careful design of systems and peripherals. However, most paradigms for computation that claim to overcome the *kTln2* limit are at least equally challenging.

The Single Spin Logic idea also has some obvious drawbacks. The generators for the ac and dc magnetic fields are necessarily bulky and therefore this paradigm is not suitable for portable electronics such as laptops and cell phones. It is more suitable for desktops and mainframe platforms. The dc magnetic field may be eliminated by the use of permanent ferromagnetic contacts as illustrated in Fig. 1, as long as the field is of the order of ~ 1 Tesla. This is what is required for InAs dots. However, the ac magnetic field still requires microwave generators which are bulky and not portable.

In the end, the advantage of Single Spin Logic is that it may be possible to increase device density 1000-fold (and therefore processing power 1000-fold) without any increase in power density on the chip. This is where, I believe, spintronics can carve out a niche.

**Acknowledgement**

This work is supported by the US Air Force Office of Scientific Research under grant FA9550-04-1-0261. I am indebted to Prof. Marc Cahay of the University of Cincinnati, Prof. Michael Forshaw of the University College, London, Prof. Vladimir Privman of Clarkson University, and Dr. Suman Datta of Intel Corporation for discussions.